\documentclass[aps,prd,epsfig,preprint,amsmath,amssymb,superscriptaddress,showpacs,nofootinbib]{revtex4}
\usepackage{multirow}
\usepackage{graphicx, bm}
\usepackage{float}
\usepackage{slashed}
\usepackage[usenames]{color}

\begin{document}

\draft

\title{Search for the anomalous $ZZZ$ and $ZZ\gamma$ gauge couplings through the process $e^+e^- \to ZZ $ with unpolarized and polarized beams}

\author{V. Cetinkaya\footnote{volkan.cetinkaya@dpu.edu.tr}}
\affiliation{\small Department of Physics, Kutahya Dumlupinar University, 43100, Turkiye.\\}

\author{S. Spor\footnote{serdar.spor@beun.edu.tr}}
\affiliation{\small Department of Medical Imaging Techniques,
Zonguldak Bülent Ecevit University, 67100, Zonguldak, Turkiye.\\}

\author{E. Gurkanli\footnote{egurkanli@sinop.edu.tr}}
\affiliation{\small Department of Physics, Sinop University, Turkiye.\\}

\author{M. K\"{o}ksal\footnote{mkoksal@cumhuriyet.edu.tr}}
\affiliation{\small Department of Physics, Sivas Cumhuriyet University, 58140, Sivas, Turkiye.}

\date{\today}

\begin{abstract}

This work offers the constraints on the anomalous neutral triple gauge couplings for the process $e^+e^- \to ZZ $ at the CLIC with $\sqrt{s}=3$ TeV. The realistic CLIC detector environments and their effects are considered in our analysis. The study is planned for the decays of produced $Z$ bosons to a pair of charged leptons (electrons or muons) and neutrino pairs. The bounds on the anomalous neutral triple gauge couplings defining $CP$-violating $C_{\widetilde{B}W}/{\Lambda^4}$ coupling and three $CP$-conserving $C_{WW}/{\Lambda^4}$, $C_{BW}/{\Lambda^4}$, and $C_{BB}/{\Lambda^4}$ couplings are obtained. Also, the effects and advantages of polarization for incoming electron beams in these calculations are investigated.

\end{abstract}


\vspace{5mm}

\maketitle


\section{Introduction}

Gauge boson pair productions in the examined processes has a significant role in the tests of the non-Abelian structure of the Standard Model (SM) and exploring physics beyond the SM. Thus, $Z\gamma$ and $ZZ$ diboson productions  can be studied with the anomalous neutral triple gauge couplings (aNTGCs) originating from high-dimensional operators defined by the effective Lagrangian method in a model-independent approach. However, $ZZ$ production from diboson productions has the smallest cross-section among all diboson processes but is highly competitive for measurements and searches due to the generally high signal-to-background ratio with leptonic or neutrino decay channels \cite{Aaboud:2019bza}. Moreover, the massive electroweak $Z$ boson has a key role in vector-boson scattering (VBS) processes for the phenomenology of electroweak interactions \cite{Beyer:2006wsa,Fleper:2017tmq}. 

The CLIC is a potential future collider that stands out due to its ability to test new physics scenarios through high-energy $e^{-}e^{+}$ collisions in a clean experimental environment with low background compared to $pp$ colliders. It also provides the opportunity for detailed study of the various processes and production mechanisms that occur in $e^{-}e^{+}$ colliders. The CLIC will be operated at three energy stages, center-of-mass energies of $\sqrt{s}=$380 GeV, 1.5 TeV and 3 TeV  \cite{Demirci:2022acw,Boland:2016yqz}, with an integrated luminosity of 5 ab$^{-1}$ at $\sqrt{s}=$3 TeV stage. CLIC is planned with $\pm 80\%$ polarization for the electron beam \cite{Roloff:2018tvb}. The weight of each event is calculated by dividing the integrated luminosity between negative and positive polarizations of the electron beam in a 4:1 ratio, resulting in integrated luminosities of $\mathcal{P}(e^-)=-80\%$ for ${\cal L}_{\text{int}}=4$ ab$^{-1}$ and $\mathcal{P}(e^-)=+80\%$ for ${\cal L}_{\text{int}}=1$ ab$^{-1}$ \cite{Roloff:2018tvb}. Beam polarizations offer advantages in future $e^+e^-$ colliders, including enhanced signals, reduced backgrounds, improved control of systematic uncertainties, and the ability to study the chiral properties of SM and BSM processes through additional observables.

On the  other hand, CLIC experiment program provides an opportunity for $\pm\%80$  polarized electron beams and no positron polarization at the center-of-mass energy of 3 TeV with the integrated luminosity of 1 ab$^{-1}$ and 4 ab$^{-1}$, respectively. Here, polarized electron beams can play a crucial role in increasing the capability of analysis. It allows a more selective probing of the process by enhancing the deviations from the SM via anomalous couplings and also the signal-background ratio. Eventually, it provides to make stringent tests and to get better constraints on anomalous couplings. 

The cross-section for a process in an electron-positron collider, utilizing polarized electron beam $\mathcal{P}(e^-)$ and polarized positron beam $\mathcal{P}(e^+)$, is determined by considering the four potential chiral cross-sections given in the following equation \cite{Fujii:2018ujn}:

\begin{eqnarray}
\label{eq.1} 
\sigma(\mathcal{P}_{e^-},\mathcal{P}_{e^+})=& \dfrac{1}{4} \Big\{(1+\mathcal{P}_{e^-})(1+\mathcal{P}_{e^+})\sigma_{RR}+(1-\mathcal{P}_{e^-})(1-\mathcal{P}_{e^+})\sigma_{LL} \\ \nonumber
&+(1+\mathcal{P}_{e^-})(1-\mathcal{P}_{e^+})\sigma_{RL}+(1-\mathcal{P}_{e^-})(1+\mathcal{P}_{e^+})\sigma_{LR}\Big\},
\end{eqnarray}

{\raggedright Here  $\sigma_{LR}$ denotes the cross-section for a left-handed polarized electron beam and a right-handed polarized positron beam. Similarly, the cross-sections $\sigma_{RL}$, $\sigma_{LL}$, and $\sigma_{RR}$ are defined following the same convention. The unpolarized cross-section is expressed as}

\begin{eqnarray}
\label{eq.2} 
\sigma_0=\frac{1}{4} \{\sigma_{RR}+\sigma_{LL}+\sigma_{RL}+\sigma_{LR}\}.
\end{eqnarray}

While the Cabibbo-Kobayashi-Maskawa (CKM) mixing matrix can explain $CP$-violation, the SM does not account for sufficient $CP$-violation effects to explain the current matter-antimatter asymmetry in the universe, resulting in the baryogenesis problem \cite{Shaposhnikovab:1987vbn,Nelson:1992ded}. Therefore, there is a need for new sources of $CP$-violation, one of Sakharov's three conditions, to explain the baryon asymmetry of the universe \cite{Sakharov:1967hla}. Investigating the $CP$-conserving and $CP$-violating couplings in the anomalous $Z\gamma\gamma$, $ZZ\gamma$ and $ZZZ$ vertex has the potential to reveal a new source of $CP$-violation and thus increases motivation to study the neutral trilinear gauge couplings.

New physics effects beyond the SM can be characterized in terms of the SM effective field theory (SMEFT), which is used to investigate aNTGC within the SM gauge group. This involves adding high-dimensional operators that are invariant under $SU(2)_L \times U(1)_Y$ to the SM Lagrangian, and determining anomalous couplings from the effective Lagrangian obtained after electroweak symmetry breaking. Processes involving neutral gauge boson couplings can be sensitive to dimension-8 operators, which are not contributed by dimension-6 operators \cite{Degrande:2014ydn}. Thus, the effective Lagrangian that accounts for SM interactions and the effects of physics beyond the SM is given by \cite{Degrande:2014ydn}

\begin{eqnarray}
\label{eq.3}
{\cal L}^{\text{NTGC}}={\cal L}_{\text{SM}}+\sum_{i}\frac{C_i}{\Lambda^{4}}({\cal O}_i+{\cal O}_i^\dagger)
\end{eqnarray}

{\raggedright where $\Lambda$ represents the new physics scale. On the other hand, $C_i$ coefficients  are dimensionless parameters and ${\cal O}_i$ dimension-8 operators  are defined by}

\begin{eqnarray}
\label{eq.4}
{\cal O}_{\widetilde{B}W}=iH^{\dagger} \widetilde{B}_{\sigma\rho}W^{\sigma\nu} \{D_\nu,D^\rho \}H,
\end{eqnarray}
\begin{eqnarray}
\label{eq.5}
{\cal O}_{BW}=iH^\dagger B_{\sigma\rho}W^{\sigma\nu} \{D_\nu,D^\rho \}H,
\end{eqnarray}
\begin{eqnarray}
\label{eq.6}
{\cal O}_{WW}=iH^\dagger W_{\sigma\rho}W^{\sigma\nu} \{D_\nu,D^\rho \}H,
\end{eqnarray}
\begin{eqnarray}
\label{eq.7}
{\cal O}_{BB}=iH^\dagger B_{\sigma\rho}B^{\sigma\nu} \{D_\nu,D^\rho \}H
\end{eqnarray}

{\raggedright where}

\begin{eqnarray}
\label{eq.8}
B_{\sigma\rho}=\left(\partial_\sigma B_\rho - \partial_\rho B_\sigma\right),
\end{eqnarray}
\begin{eqnarray}
\label{eq.9}
W_{\sigma\rho}=\sigma^i\left(\partial_\sigma W_\rho^i - \partial_\rho W_\sigma^i + g\epsilon_{ijk}W_\sigma^j W_\rho^k\right),
\end{eqnarray}

{\raggedright with $\langle \sigma^i\sigma^j\rangle=\delta^{ij}/2$ and}

\begin{eqnarray}
\label{eq.10}
D_\mu \equiv \partial_\mu - i\frac{g^\prime}{2}B_\mu Y - ig_W W_\mu^i\sigma^i.
\end{eqnarray}

{\raggedright Here, $H$ is the Higgs field, $B_{\sigma \rho}$ and $W_{\sigma \rho}$ are the field strength tensors and $D_{\mu}$ shows the covariant derivative.}

For $ZZ$ production at high-energy scale, the most significant contribution of physics beyond the SM is provided by the interference between the SM and the dimension-8 operators. Unless interference between the SM and dimension-8 and dimension-10 operators are both influentially suppressed, the ${\cal O}({\Lambda^{-8}})$ square of dimension-8 operators cannot be expected to contain a contribution from physics beyond the SM when the energy scale is high. At the tree-level, while dimension-6 operators have no effect on the anomalous neutral gauge boson couplings, the contributions of dimension-8 operators are of the order ${\upsilon^2\hat{s}}/{\Lambda^4}$. Nevertheless, dimension-6 operators can have ${\upsilon^2\hat{s}}/{\Lambda^4}$ order effects on the aNTGC at one-loop. Therefore, the contribution of dim-8 operators appears to be more important than the one-loop contribution of dimension-6 operators with  $\Lambda \lesssim \sqrt{4\pi\hat{s}/\alpha}$ (Here, $\upsilon$ and $\alpha$ are vacuum expectation values and fine-structure constant) \cite{Degrande:2014ydn}.  

In the literature, effective Lagrangian for the aNTGCs with dim-6 and dim-8 operators is given by \cite{Gounaris:2000wlp}

\begin{eqnarray}
\label{eq.11}
\begin{split}
{\cal L}=&\frac{e}{m_Z^2}\Bigg[-[f_4^\gamma(\partial_\mu F^{\mu\beta})+f_4^Z(\partial_\mu Z^{\mu\beta})]Z_\alpha (\partial^\alpha Z_\beta)+[f_5^\gamma(\partial^\sigma F_{\sigma\mu})+f_5^Z (\partial^\sigma Z_{\sigma\mu})]\widetilde{Z}^{\mu\beta}Z_\beta  \\
&-[h_1^\gamma (\partial^\sigma F_{\sigma\mu})+h_1^Z (\partial^\sigma Z_{\sigma\mu})]Z_\beta F^{\mu\beta}-[h_3^\gamma(\partial_\sigma F^{\sigma\rho})+h_3^Z(\partial_\sigma Z^{\sigma\rho})]Z^\alpha \widetilde{F}_{\rho\alpha}   \\
&-\bigg\{\frac{h_2^\gamma}{m_Z^2}[\partial_\alpha \partial_\beta \partial^\rho F_{\rho\mu}]+\frac{h_2^Z}{m_Z^2}[\partial_\alpha \partial_\beta(\square+m_Z^2)Z_\mu]\bigg\}Z^\alpha F^{\mu\beta}   \\
&+\bigg\{\frac{h_4^\gamma}{2m_Z^2}[\square\partial^\sigma F^{\rho\alpha}]+\frac{h_4^Z}{2m_Z^2}[(\square+m_Z^2)\partial^\sigma Z^{\rho\alpha}]\bigg\}Z_\sigma\widetilde{F}_{\rho\alpha}\Bigg]
\end{split}
\end{eqnarray}

{\raggedright where $\widetilde{Z}_{\mu\nu}=1/2\epsilon_{\mu\nu\rho\sigma}Z^{\rho\sigma}$ $(\epsilon^{0123}=+1)$ is defined by the field strength tensor $Z_{\mu\nu}=\partial_\mu Z_\nu - \partial_\nu Z_\mu$ and similarly the electromagnetic field tensor $F_{\mu\nu}$ is also defined. Nevertheless, $f_4^V$, $h_1^V$, $h_2^V$ are the $CP$-violating couplings and $f_5^V$, $h_3^V$, $h_4^V$ are the $CP$-conserving couplings $(V=\gamma$, $Z)$. In the SM, these couplings are zero at tree-level. In Eq.~(\ref{eq.11}), $h_2^V$ and $h_4^V$ couplings are concerned with dim-8 operators while the other couplings to dim-6 operators.}

The $CP$-conserving anomalous couplings for the $ZZV^*$ vertex with the two on-shell $Z$ bosons and one off-shell $V^*=\gamma ,Z$ boson are written by \cite{Degrande:2014ydn}

\begin{eqnarray}
\label{eq.12}
f_5^Z=0,
\end{eqnarray}
\begin{eqnarray}
\label{eq.13}
f_5^\gamma=\frac{\upsilon^2 m_Z^2}{4c_\omega s_\omega} \frac{C_{\widetilde{B}W}}{\Lambda^4}.
\end{eqnarray}

Nevertheless, the $CP$-violating anomalous couplings for the $ZZV^*$ vertex are given as follows

\begin{eqnarray}
\label{eq.14}
f_4^Z=\frac{m_Z^2 \upsilon^2 \left(c_\omega^2 \frac{C_{WW}}{\Lambda^4}+2c_\omega s_\omega \frac{C_{BW}}{\Lambda^4}+4s_\omega^2 \frac{C_{BB}}{\Lambda^4}\right)}{2c_\omega s_\omega},
\end{eqnarray}
\begin{eqnarray}
\label{eq.15}
f_4^\gamma=-\frac{m_Z^2 \upsilon^2 \left(-c_\omega s_\omega \frac{C_{WW}}{\Lambda^4}+\frac{C_{BW}}{\Lambda^4}(c_\omega^2-s_\omega^2)+4c_\omega s_\omega \frac{C_{BB}}{\Lambda^4}\right)}{4c_\omega s_\omega},
\end{eqnarray}

{\raggedright where $s_\omega$ and $c_\omega$ are the sine and cosine of weak mixing angles $\theta_{w}$.}

On the theoretical side, contributions to NTGC in the context of the minimal supersymmetric standard model (MSSM) \cite{Gounaris:2000jsx,Choudhury:2001yvz,Gounaris:2006edf}, little Higgs model \cite{Dutta:2009tvq}, and two-Higgs doublet model (THDM) \cite{Chang:1995gwx,Grzadkowski:2016evb,Belusca:2018hhd} have been discussed previously. There has recently been extensive research on the potential effects of NTGC in collider phenomenology, for example, $e^-e^+$ collider \cite{Hagiwara:1987fqw,Rodriguez:2009rnw,Ananthanarayan:2014cal,Rahaman:2016nzs,Rahaman:2017qed,Ellis:2020ekm,Fu:2021jec,Ellis:2021rop,Spor:2022ssp,Yang:2022tgw,Jahedi:2023abc}, $pp$ collider \cite{Senol:2018gvg,Rahaman:2019tnp,Senol:2019ybv,Senol:2020hbh,Yilmaz:2020ser,Yilmaz:2021dbm,Hernandez:2021wsz,Lombardi:2022tgv} and $\mu^-\mu^+$ collider \cite{Senol:2022psb,Spor:2023ywc}.

The current experimental limits on $C_{BB}/{\Lambda^4}$, $C_{BW}/{\Lambda^4}$, $C_{\widetilde{B}W}/{\Lambda^4}$ and $C_{WW}/{\Lambda^4}$ couplings at the $95\%$ Confidence Level (C.L.) by  ATLAS and CMS Collaborations at the LHC with center-of-mass energy of 13 TeV are given in Table~\ref{tab1}.

\begin{table}[H]
\caption{The experimental limits at 95$\%$ C.L. on anomalous $C_{BB}/{\Lambda^4}$, $C_{BW}/{\Lambda^4}$, $C_{\widetilde{B}W}/{\Lambda^4}$ and $C_{WW}/{\Lambda^4}$ couplings.}
\label{tab1}
\begin{ruledtabular}
\begin{tabular}{lcccc}
\multirow{2}{*} & \multicolumn{4}{c}{Couplings (TeV$^{-4}$)}\\
 & ${C_{BB}}/{\Lambda^4}$ & ${C_{BW}}/{\Lambda^4}$ & ${C_{\widetilde{B}W}}/{\Lambda^4}$ & ${C_{WW}}/{\Lambda^4}$\\ \hline
ATLAS \cite{Aaboud:2018ybz} $Z\gamma\rightarrow\nu\bar{\nu}\gamma$ & \multirow{2}{*}{[-0.24; 0.24]} & \multirow{2}{*}{[-0.65; 0.64]} & \multirow{2}{*}{[-1.10; 1.10]} & \multirow{2}{*}{[-2.30; 2.30]}\\ 
($\sqrt{s}=13$ TeV, ${\cal L}_{\text{int}}=36.1$ fb$^{-1}$) & & & & \\ \hline

ATLAS \cite{Aaboud:2018onm} $ZZ\rightarrow\ell^+\ell^-\ell^{\prime+}\ell^{\prime-}$ & \multirow{2}{*}{[-2.70; 2.80]} & \multirow{2}{*}{[-3.30; 3.30]} & \multirow{2}{*}{[-5.90; 5.90]} & \multirow{2}{*}{[-3.00; 3.00]}\\ 
($\sqrt{s}=13$ TeV, ${\cal L}_{\text{int}}=36.1$ fb$^{-1}$) & & & & \\ \hline

CMS \cite{Sirunyan:2021edk} $ZZ\rightarrow\ell^+\ell^-\ell^{\prime+}\ell^{\prime-}$ & \multirow{2}{*}{[-1.20; 1.20]} & \multirow{2}{*}{[-1.40; 1.30]} & \multirow{2}{*}{[-2.30; 2.50]} & \multirow{2}{*}{[-1.40; 1.20]}\\ 
($\sqrt{s}=13$ TeV, ${\cal L}_{\text{int}}=137$ fb$^{-1}$) & & & & \\
\end{tabular}
\end{ruledtabular}
\end{table}

\section{Generation of events}

\begin{figure}[H]
\centerline{\scalebox{0.75}{\includegraphics{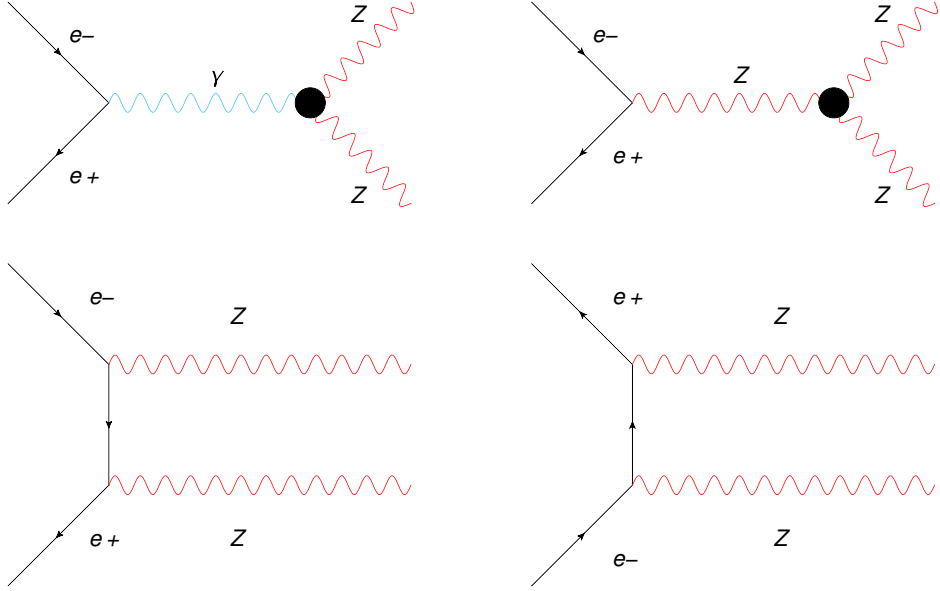}}}
\caption{Schematic diagrams for the process $e^- e^{+} \to ZZ $ including the anomalous contribution of $ZZZ$ and $ZZ\gamma$ vertices and the SM.}
\label{Fig.1}
\end{figure}

Our aim in the SMEFT framework is to determine the constraints on the aNTGCs in $ZZ$ diboson production at the CLIC. We present the tree-level Feynman diagrams for the process $e^+e^- \to ZZ$ in Fig.~\ref{Fig.1}, where black dots indicate the anomalous $ZZ\gamma$ and $ZZZ$ couplings representing new physics, while the two other diagrams show the SM contributions.

We have conducted the production of signal events through the process $e^+e^-\rightarrow ZZ\rightarrow \ell \ell \nu\nu$ and background events at the CLIC using the Universal FeynRules Output (UFO) model file \cite{Degrande:2014ydn} with MadGraph5$\_$aMC@NLO \cite{Alwall:2014cvc}. For each coupling, 500k signal and background events were generated. We utilized the CLIC detector card in Delphes v3.5.0 \cite{Favereau:2014qaz} to include realistic detector effects, and applied PYTHIA v8.306 \cite{Bierlich:2022tge} for parton showering and hadronization. The generated events were analyzed with the ExRootAnalysis \cite{ExRootAnalysis} package and ROOT v6.26 \cite{Brun:1997cvb}. 

\begin{figure}[h!]
\includegraphics[scale=0.75]{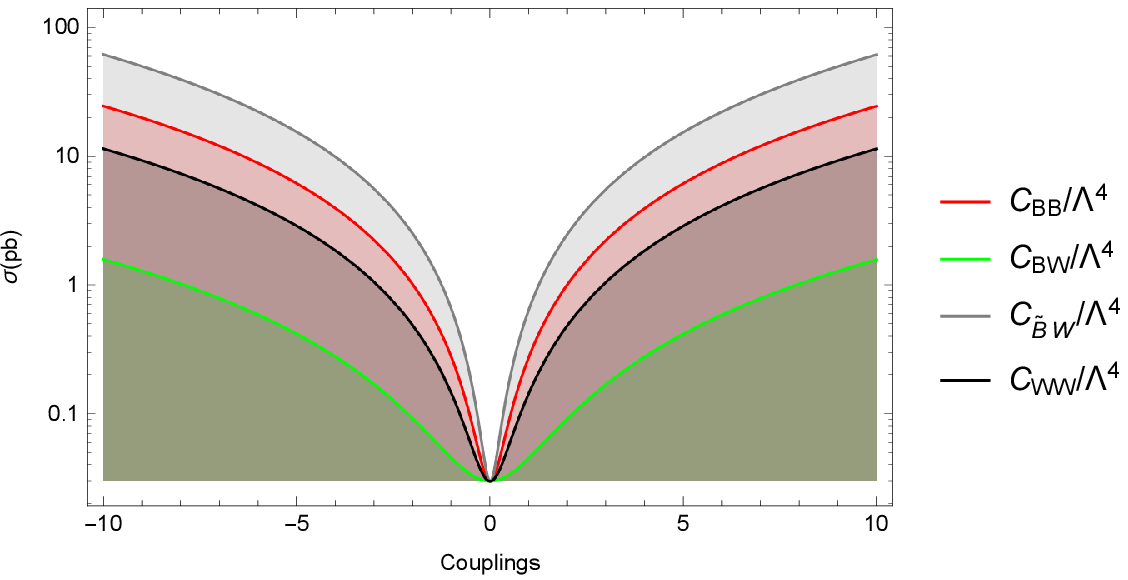}\includegraphics[scale=0.75]{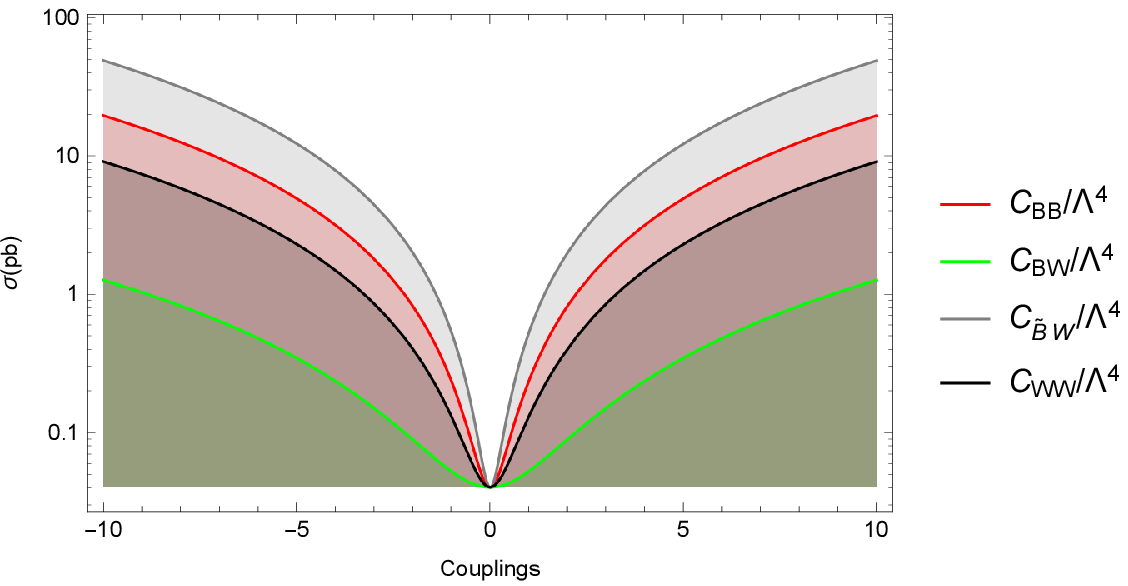}\\
\caption{Cross-section of $e^{-} e^{+} \to ZZ $ process in terms of the anomalous $C_{BB}/{\Lambda^4}$, $C_{BW}/{\Lambda^4}$, $C_{\widetilde{B}W}/{\Lambda^4}$, $C_{WW}/{\Lambda^4}$ couplings at the $\sqrt{s}=3$ TeV and $P_{e^-}=-80\%$. Here, in the figure on the right hand side, ISR and beamstrahlung effects are taken into account.}
\label{Fig.2}
\end{figure}

\begin{figure}[h!]
\includegraphics[scale=0.75]{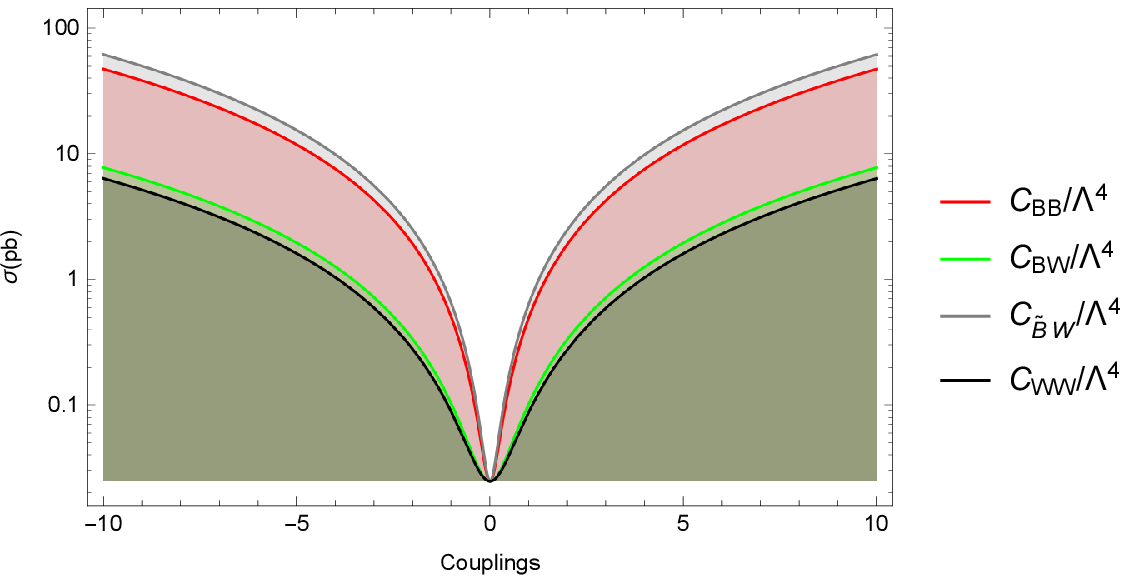}\includegraphics[scale=0.75]{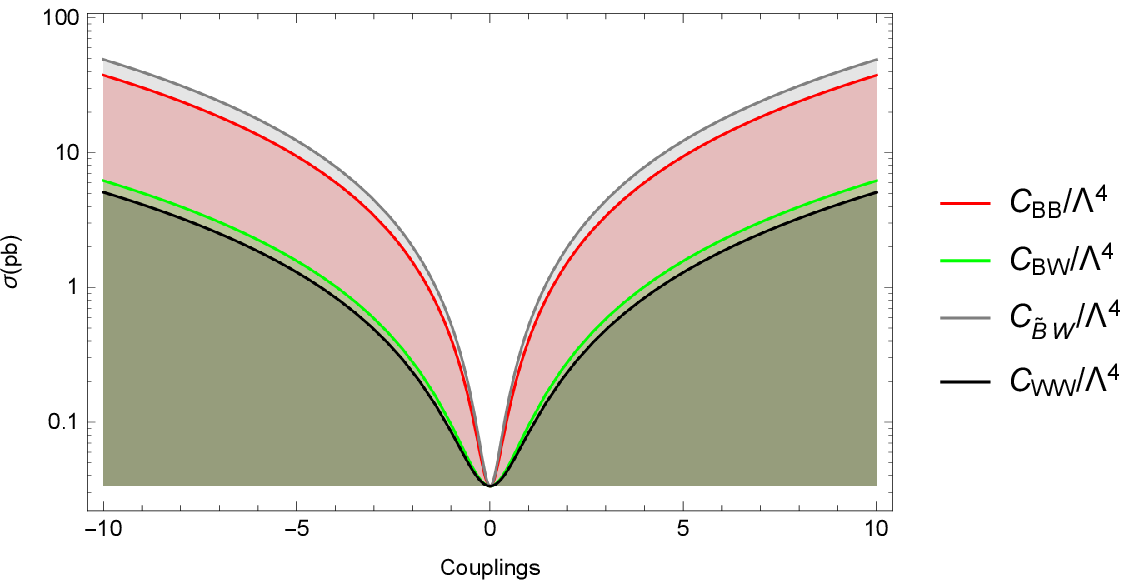}\\
\caption{Same as Fig.~\ref{Fig.2} but for unpolarized beams $P_{e^-}=0\%$.}
\label{Fig.3}
\end{figure}

\begin{figure}[h!]
\includegraphics[scale=0.75]{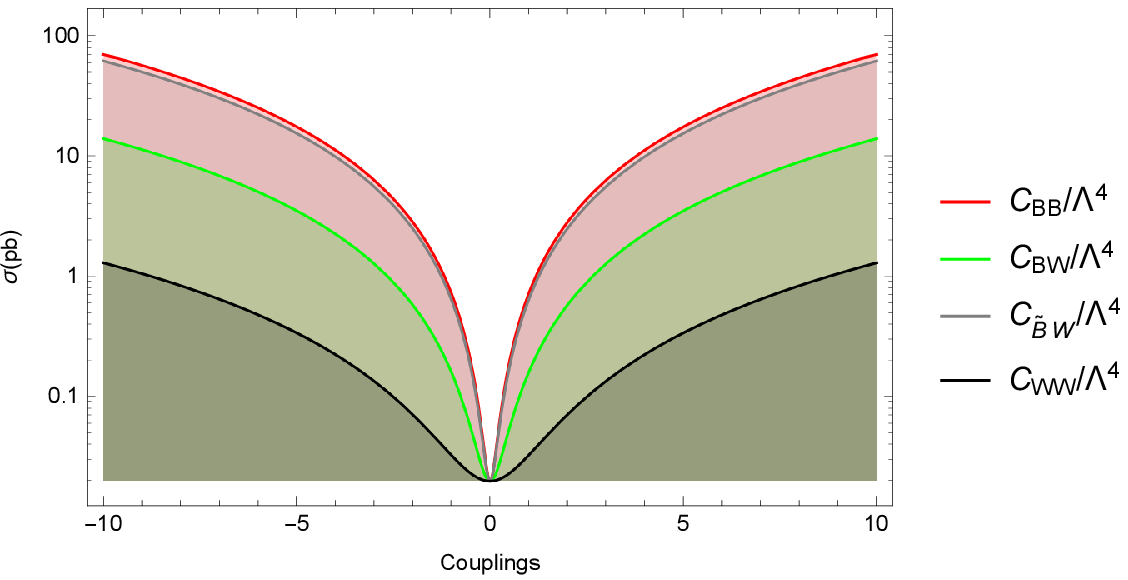}\includegraphics[scale=0.75]{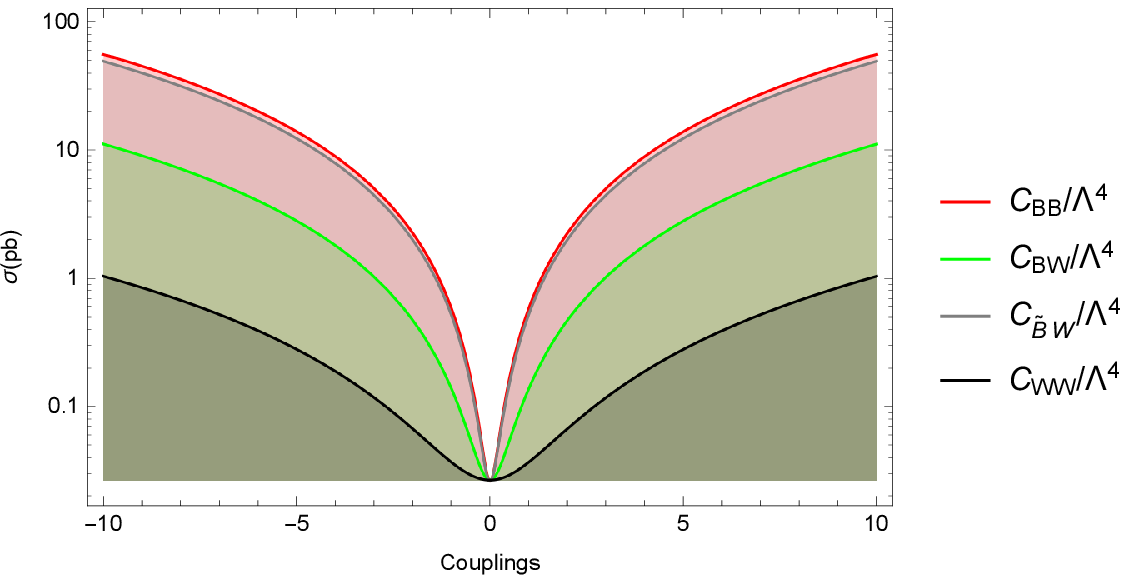}\\
\caption{Same as Fig.~\ref{Fig.2} but for polarized beams $P_{e^-}=80\%$.}
\label{Fig.4}
\end{figure}

The total cross-sections of the $e^-e^+\rightarrow ZZ$ process with and without the ISR and beamstrahlung effects at $\sqrt{s}=3$ TeV at the CLIC are given in Figs.~\ref{Fig.2}-\ref{Fig.4} for $-80\%$, unpolarized, and $+80\%$ electron beam polarizations, respectively. The cross-sections are plotted as a function of the anomalous couplings $C_{BB}/{\Lambda^4}$, $C_{BW}/{\Lambda^4}$, $C_{\widetilde{B}W}/{\Lambda^4}$, and $C_{WW}/{\Lambda^4}$. Here, only one coupling is run at a time while the other three couplings were set to zero. However, the cross-sections of the $e^-e^+\rightarrow ZZ$ process with and without the ISR and beamstrahlung effects for $C_{BB}/{\Lambda^4}=3$ TeV$^{-4}$, $C_{BW}/{\Lambda^4}=3$ TeV$^{-4}$, $C_{\widetilde{B}W}/{\Lambda^4}=3$ TeV$^{-4}$, and $C_{WW}/{\Lambda^4}=3$ TeV$^{-4}$ are given numerically in Table~\ref{tab2}. In all three polarization cases, the cross-sections of the signal are effectively reduced in the presence of ISR and beamstrahlung effects compared to their absence, and the percentage changes are shown in Table~\ref{tab2}.

\begin{table}[H]
\centering
\caption{Cross sections (pb) of the process $e^{-} e^{+} \to ZZ$ with and without the ISR and beamstrahlung effects according to $P_{e^-}=0\%, -80\%, 80\%$ at the 3 TeV CLIC.}
\label{tab2}
\centering
\begin{tabular}{p{2.6cm}p{1.3cm}p{1.5cm}p{1.2cm}p{0.1cm}p{1.3cm}p{1.5cm}p{1.2cm}p{0.1cm}p{1.3cm}p{1.5cm}p{1.2cm}}
\hline \hline
Couplings & \multicolumn{3}{c}{Unpolarized} && \multicolumn{3}{c}{$P_{e^-}=-80\%$} && \multicolumn{3}{c}{$P_{e^-}=80\%$}\\ \cline{2-4} \cline{6-8} \cline{10-12}
(TeV$^{-4}$) & no ISR & with ISR & && no ISR & with ISR & && no ISR & with ISR & \\ \hline
$C_{BB}/{\Lambda^4}=3$ & 4.277 & 3.434 & -19.7\% && 2.240 & 1.806 & -19.3\% && 6.308 & 5.033 & -20.2\%\\ 
$C_{BW}/{\Lambda^4}=3$ & 0.722 & 0.589 & -18.4\% && 0.169 & 0.151 & -10.6\% && 1.274 & 1.029 & -19.2\%\\
$C_{\widetilde{B}W}/{\Lambda^4}=3$ & 5.576 & 4.457 & -20.0\% && 5.572 & 4.463 & -19.9\% && 5.582 & 4.453 & -20.2\%\\ 
$C_{WW}/{\Lambda^4}=3$ & 0.596 & 0.489 & -17.9\% && 1.058 & 0.861 & -18.6\% && 0.134 & 0.117 & -12.6\%\\    \hline \hline
\end{tabular}
\end{table}

The final state $\ell \ell \nu\nu$ of the process $e^+e^-\rightarrow ZZ$ consists of a pair of charged leptons and neutrinos, coming from the decay of $Z$ bosons. This process with non-zero effective couplings is assumed as signal including SM contribution and interference term that is coming from crossing the diagrams of effective couplings and SM. The main source of SM background process considered in this paper is the $e^+e^-\rightarrow \ell \ell \nu\nu$. On the other hand, ISR and beamstrahlung effects also give considerable contributions to the background in lepton colliders. These effects were not considered because of the incompatibility of the simulation tools used in the study. In this case, the sensitivities on aNTGC were slightly optimistic compared with a more comprehensive study and experiment.

\begin{figure}[H]
\centerline{\scalebox{0.80}{\includegraphics{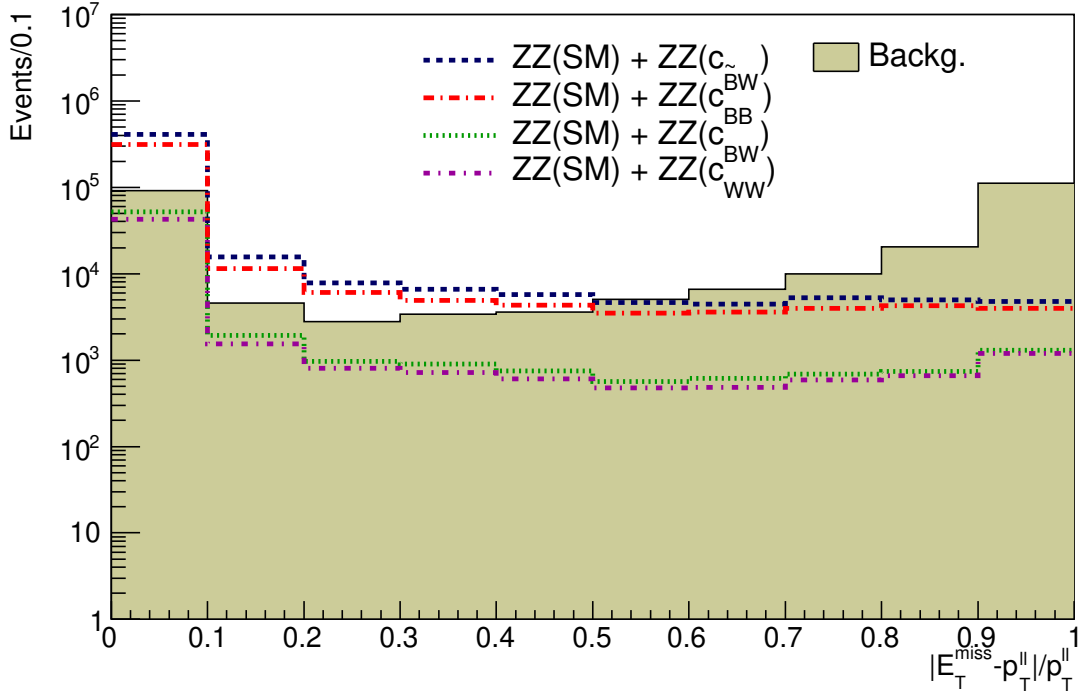}}}
\caption{Distribution of transverse momentum balance ratio for signal with $C_{\widetilde{B}W}/{\Lambda^4}=3$ TeV$^{-4}$, $C_{BB}/{\Lambda^4}=3$ TeV$^{-4}$, $C_{BW}/{\Lambda^4}=3$ TeV$^{-4}$, $C_{WW}/{\Lambda^4}=3$ TeV$^{-4}$ and relevant background processes}
\label{Fig.5}
\end{figure}

The preselection criteria for the analysis include the presence of a dilepton of the same flavor with opposite charge to form the $Z$ boson. At least two leptons of the same flavor with opposite charge ($N_{\ell\,(e,\,\mu)} >= 2$) must be present in the event. The transverse momentum of the leading lepton $\ell_{1}$ must be greater than 30 GeV and the sub-leading lepton $\ell_{2}$ greater than 20 GeV. A cut on the pseudorapidity between the two leptons is also imposed with a limit of $|\eta| < 2.5$. As can be seen from Fig.~\ref{Fig.5}, the event selection is optimized by imposing $|E_T^{miss}-p_T^{\ell\ell}|/p_T^{\ell\ell} < 0.5$, as the distribution of transverse momentum balance ratio for the signal deviates significantly from the SM background for values less than 0.5. 

\begin{figure}[H]
\centerline{\scalebox{0.8}{\includegraphics{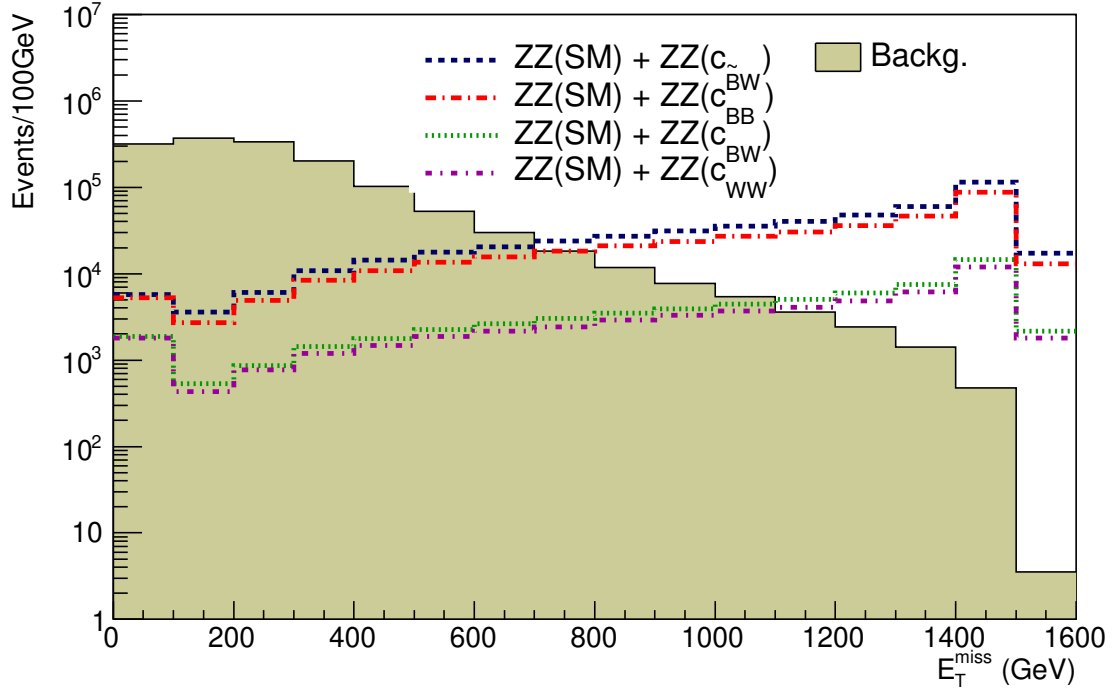}}}
\caption{Distribution of missing energy transverse for signal with $C_{\widetilde{B}W}/{\Lambda^4}=3$ TeV$^{-4}$, $C_{BB}/{\Lambda^4}=3$ TeV$^{-4}$, $C_{BW}/{\Lambda^4}=3$ TeV$^{-4}$, $C_{WW}/{\Lambda^4}=3$ TeV$^{-4}$ and relevant background processes}
\label{Fig.6}
\end{figure}

\begin{figure}[H]
\centerline{\scalebox{0.8}{\includegraphics{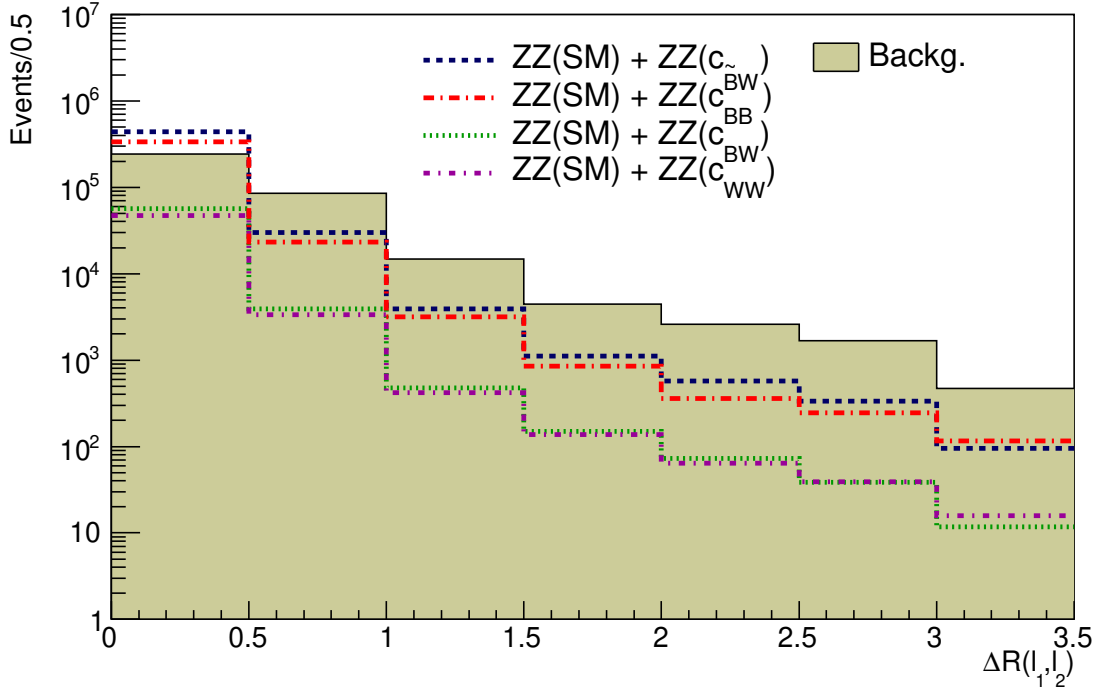}}}
\caption{Distance $\Delta R({\ell_1,\ell_2})$ between leading and sub-leading charged leptons for signal with $C_{\widetilde{B}W}/{\Lambda^4}=3$ TeV$^{-4}$, $C_{BB}/{\Lambda^4}=3$ TeV$^{-4}$, $C_{BW}/{\Lambda^4}=3$ TeV$^{-4}$, $C_{WW}/{\Lambda^4}=3$ TeV$^{-4}$ and relevant background processes}
\label{Fig.7}
\end{figure}

As seen in Fig.~\ref{Fig.6}, choosing $E_T^{miss}>400$ GeV provides a well seperation region between the signals and backgrounds. On the other hand, by limiting the distance between two charged leptons in the $\eta-\phi$ plane to $\Delta R({\ell_{1},\ell_{2}}) < 1.5$, the $\ell\ell\nu\nu$ background can be suppressed as seen in Fig.~\ref{Fig.7}. Additionally, the events must not have any extra leptons with a transverse momentum greater than 10 GeV.

\begin{figure}[H]
\centerline{\scalebox{0.8}{\includegraphics{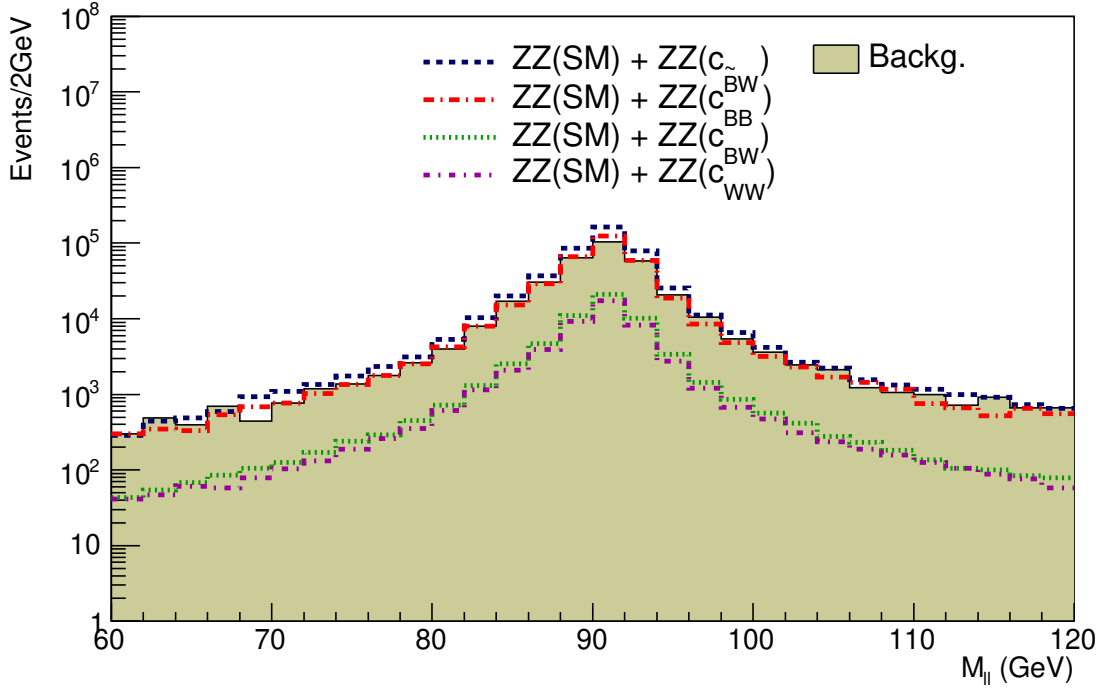}}}
\caption{Invariant mass distribution of $Z(\ell\ell)$ boson for signal with $C_{\widetilde{B}W}/{\Lambda^4}=3$ TeV$^{-4}$, $C_{BB}/{\Lambda^4}=3$ TeV$^{-4}$, $C_{BW}/{\Lambda^4}=3$ TeV$^{-4}$, $C_{WW}/{\Lambda^4}=3$ TeV$^{-4}$ and relevant background processes}
\label{Fig.8}
\end{figure}

\begin{figure}[H]
\centerline{\scalebox{0.8}{\includegraphics{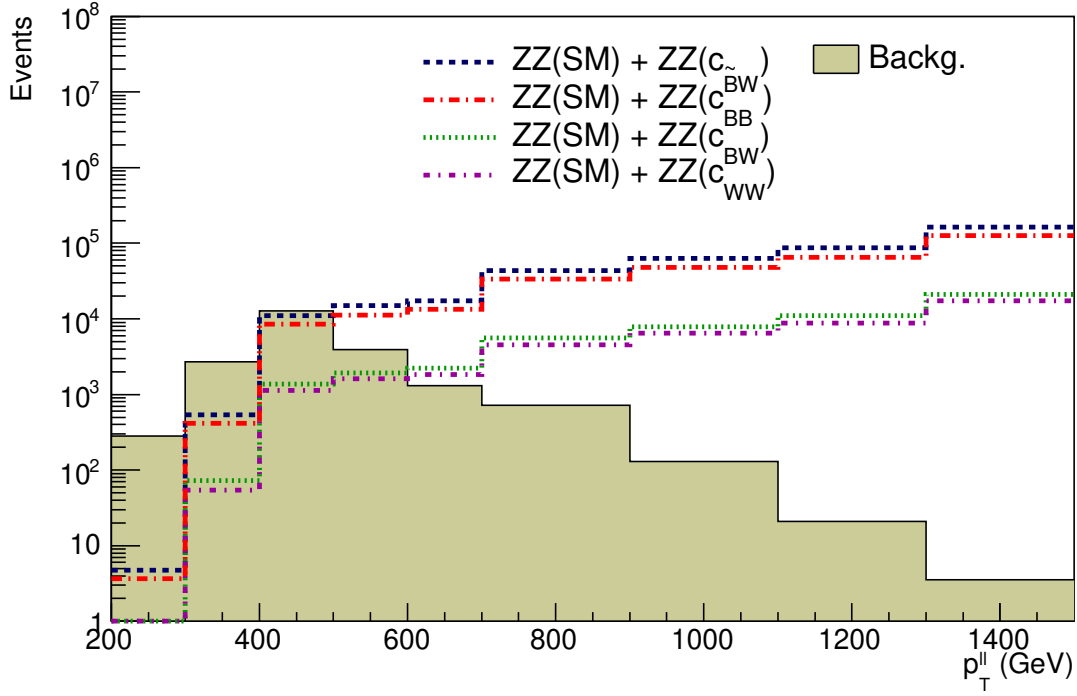}}}
\caption{Distribution of the transverse momentum of dilepton $p^{\ell \ell}_{T}$ for signal with $C_{\widetilde{B}W}/{\Lambda^4}=3$ TeV$^{-4}$, $C_{BB}/{\Lambda^4}=3$ TeV$^{-4}$, $C_{BW}/{\Lambda^4}=3$ TeV$^{-4}$, $C_{WW}/{\Lambda^4}=3$ TeV$^{-4}$ and relevant background processes}
\label{Fig.9}
\end{figure}

If the mass of the $Z$ boson is $M_Z$, we determine the invariant mass to be in the range of $|M_{\ell\ell}-M_Z| < 20$ GeV to focus on events from the decay of the two charged leptons seen in Fig.~\ref{Fig.8}. The kinematic distributions utilized to distinguish the signal from the background in this study are shown in Figs.~\ref{Fig.5}-\ref{Fig.8}. Each plot is created by applying generator level cuts to determine the optimum cut set to separate the signals and background events. All selected cuts are given in a cut-flow chart for the analysis in Table~\ref{tab3}. A kinematic cut, such as the invisible invariant mass $M_{\nu\nu}$, can eliminate cross-contamination by separating the signal from the background. Kinematic cuts have been minimally used to simplify this study and to provide a more general perspective on the design parameters of the future lepton collider, the CLIC. For this reason, we have not considered the invisible invariant mass, but it is evident that the results will be slightly worse if the invisible invariant mass is not used. Finally, distributions of the transverse momentum of dilepton system $p^{\ell \ell}_{T}$ ($\vec{p}_{T}^{\,\,\ell \ell}=\vec{p}_{T}^{\,\,\ell_1}+\vec{p}_{T}^{\,\,\ell_2}$) after applied cuts in Table~\ref{tab3} for every single coupling is given in Fig. \ref{Fig.9}. While obtaining the sensitivities of the anomalous $C_{\widetilde{B}W}/{\Lambda^4}$, $C_{WW}/{\Lambda^4}$, $C_{BW}/{\Lambda^4}$, and $C_{BB}/{\Lambda^4}$ couplings, we have focused on the variable transverse momentum of the dilepton system.

\begin{table} [H]
\centering
\caption{List of selected cuts for the analysis.}
\label{tab3}
\begin{tabular}{p{3cm}p{13cm}}
\hline \hline
Cuts & Definitions \\ 
\hline
Cut-0 & Preselection: $N_{\ell\,(e,\,\mu)} >= 2$ and neutrino pairs\\ 
Cut-1 & Transverse momentum and pseudo-rapidity of the leading and sub-leading charged leptons: $p_T^{\ell_1} > 30$ GeV, $p_T^{\ell_2} > 20$ GeV and $|\eta^{\ell_1}| < 2.5$, $|\eta^{\ell_2}| < 2.5$\\ 
Cut-2 & $p_T$ balance: $|E_T^{miss}-p_T^{\ell\ell}|/p_T^{\ell\ell} < 0.5$\\ 
Cut-3 & Missing energy transverse: $E_T^{miss} > 400$ GeV\\
Cut-4 & Minimum distance between leptons: $\Delta R({\ell_{1},\ell_{2}}) < 1.5$\\
Cut-5 & Invariant mass: $|M_{\ell\ell}-M_Z| < 20$ GeV\\ \hline \hline
\end{tabular}
\end{table}

\section{Sensitivities on aNTGC}

 The sensitivities obtained at 95$\%$ C. L. on the anomalous $C_{BB}/{\Lambda^4}$, $C_{BW}/{\Lambda^4}$, $C_{\widetilde{B}W}/{\Lambda^4}$, and $C_{WW}/{\Lambda^4}$ couplings are calculated with the process of $e^-e^+\,\rightarrow\,ZZ$ for the decay for lepton and neutrino pairs of $Z$ bosons using $\chi^2$ method for every single bin given in the following equation:

\begin{eqnarray}
\label{eq.16}
\chi^{2}=\sum_{i}^{n_{bins}} (\frac{N_{i}^{TOT}-N_{i}^{B}}{N_{i}^{B}\Delta_{i}})^{2}
\end{eqnarray}

{\raggedright Here, $N_{i}^{TOT}$ and $N_{i}^{B}$ are the number of events of the anomalous couplings and the SM background, respectively. On the other hand, $\Delta_{i}=\sqrt{1/N_{i}^{B}}$ is the statistical uncertainty for every bin. The limits at 95$\%$ C.L. on anomalous $C_{BB}/{\Lambda^4}$, $C_{BW}/{\Lambda^4}$, $C_{\widetilde{B}W}/{\Lambda^4}$, and $C_{WW}/{\Lambda^4}$ couplings via the process $e^-e^+\,\rightarrow\,ZZ$  with various electron polarization with the corresponding integrated luminosities due to the CLIC program are given in Table~\ref{tab4}. Our best sensitivities are given as follows}

\begin{eqnarray}
\label{eq.17} 
C_{BB}/{\Lambda^4}=[-1.73; 2.21]\times10^{-2}\,\text{TeV}^{-4}\,,
\end{eqnarray}
\begin{eqnarray}
\label{eq.18} 
C_{BW}/{\Lambda^4}=[-4.55; 4.56]\times10^{-2}\,\text{TeV}^{-4}\,,
\end{eqnarray}
\begin{eqnarray}
\label{eq.19} 
C_{\widetilde{B}W}/{\Lambda^4}=[-1.29; 1.79]\times10^{-2}\,\text{TeV}^{-4}\,,
\end{eqnarray}
\begin{eqnarray}
\label{eq.20} 
C_{WW}/{\Lambda^4}=[-6.78; 6.83]\times10^{-2}\,\text{TeV}^{-4}\,.
\end{eqnarray}

\begin{figure}[H]
\centerline{\scalebox{1.2}{\includegraphics{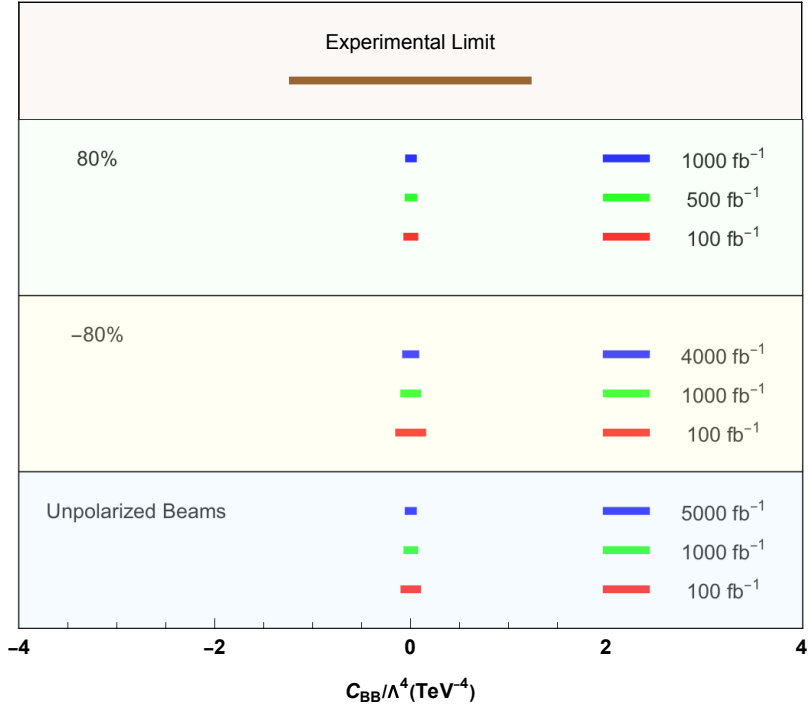}}}
\caption{Comparison of the experimental limits and obtained sensitivity on the anomalous $C_{BB}/\Lambda^4$ coupling for various luminosities for every polarization states of electron beam. }
\label{Fig.10}
\end{figure}

\begin{figure}[H]
\centerline{\scalebox{1.2}{\includegraphics{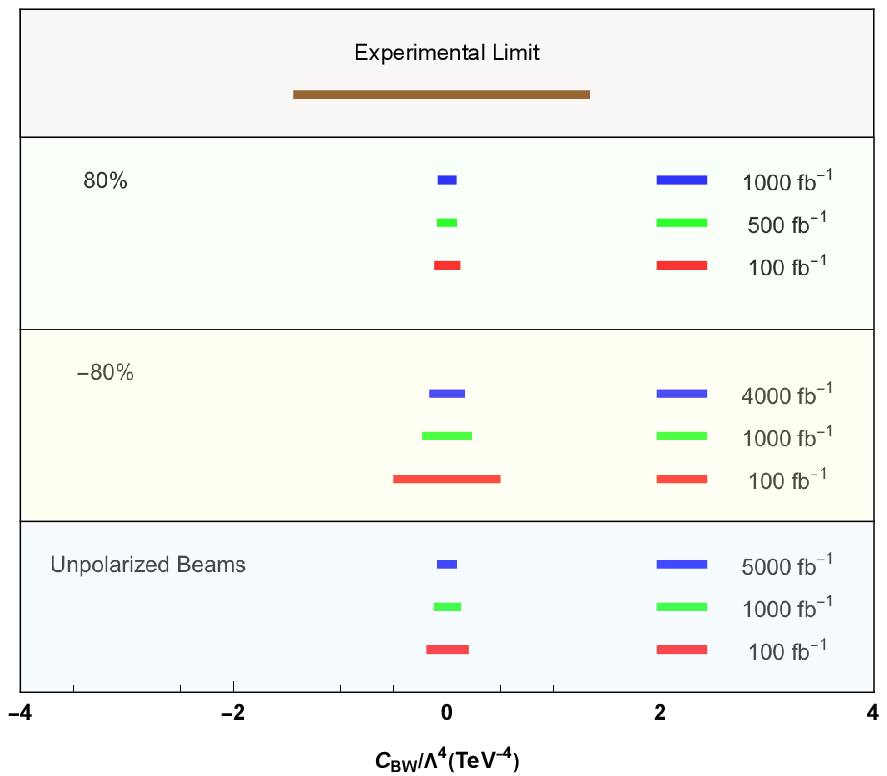}}}
\caption{Same as Fig.~\ref{Fig.10} but for the anomalous $C_{BW}/\Lambda^4$ coupling.}
\label{Fig.11}
\end{figure}

\begin{figure}[H]
\centerline{\scalebox{1.2}{\includegraphics{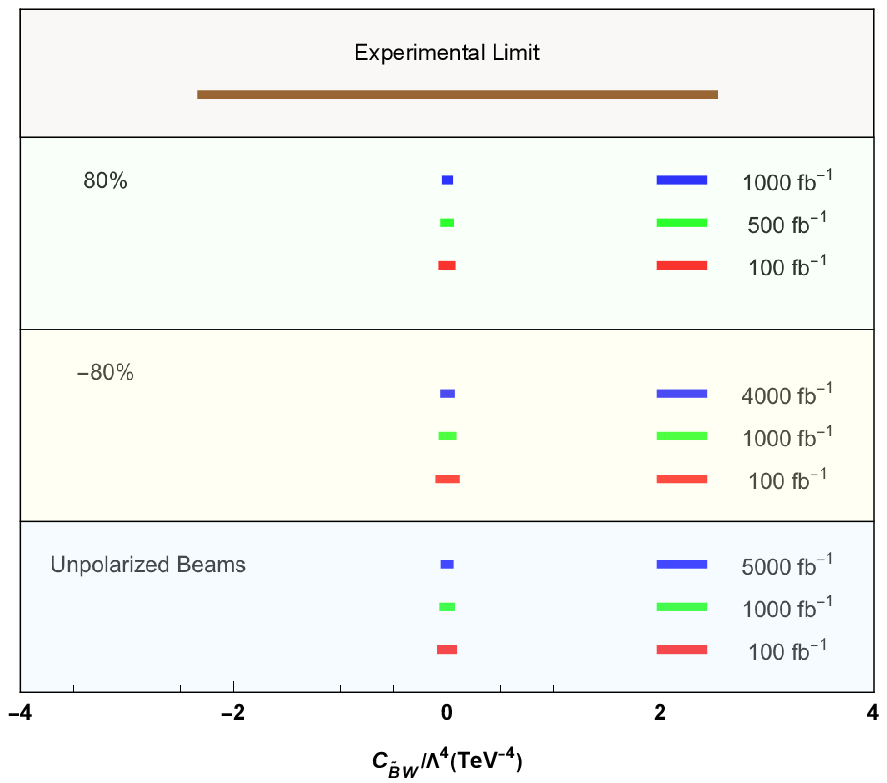}}}
\caption{Same as Fig.~\ref{Fig.10} but for the anomalous $C_{\widetilde{B}W}/\Lambda^4$ coupling.}
\label{Fig.12}
\end{figure}

\begin{figure}[H]
\centerline{\scalebox{1.2}{\includegraphics{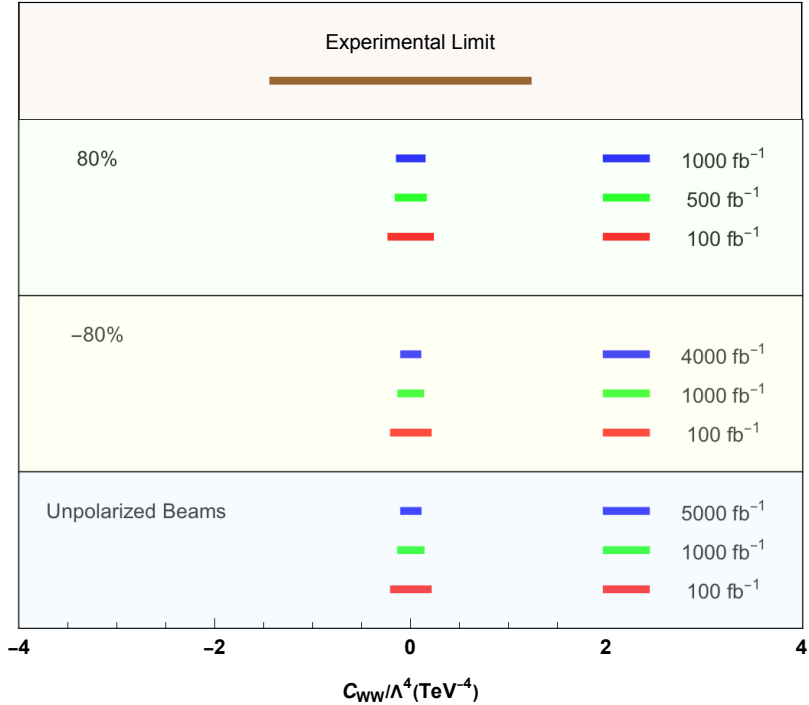}}}
\caption{Same as Fig.~\ref{Fig.10} but for the anomalous $C_{WW}/\Lambda^4$ coupling.}
\label{Fig.13}
\end{figure}

It can be seen that the most sensitive coupling is $C_{\widetilde{B}W}/{\Lambda^4}$, with a value of $[-1.29; 1.79]\times10^{-2}$ TeV$^{-4}$. The sensitivities of the anomalous $C_{\widetilde{B}W}/{\Lambda^4}$, $C_{BW}/{\Lambda^4}$ and $C_{BB}/{\Lambda^4}$ couplings is improved with $80\%$ polarized electron beams compared to unpolarized and $-80\%$ polarized beams. However, the $C_{WW}/{\Lambda^4}$ couplings have the best sensitivity with $-80\%$ polarized beams. The obtained sensitivities are compared with the latest experimental results using bar-chart graphics in Figs.~\ref{Fig.10}-\ref{Fig.13} for unpolarized and polarized beams with different integrated luminosities. Our results show that the most sensitive limits in  Eqs.~(\ref{eq.17})-(\ref{eq.20}) are approximately 18-140 times better than the experimental limits in Ref.~\cite{Sirunyan:2021edk}.

\begin{table}[H]
\caption{Sensitivities on aNTGCs $C_{BB}/\Lambda^{4}$, $C_{BW}/\Lambda^{4}$, $C_{\widetilde{B}W}/\Lambda^{4}$ and $C_{WW}/\Lambda^{4}$ through the process $e^+e^- \to ZZ$ at CLIC. }
\label{tab4}
\begin{ruledtabular}
\begin{tabular}{cccc}
$P_{e^-}$              & $0\%$       & $-80\%$    & $80\%$ \\
\hline
Couplings (TeV$^{-4}$) & ${\cal L}=5$ ab$^{-1}$ & ${\cal L}=4$ ab$^{-1}$ & ${\cal L}=1$ ab$^{-1}$ \\
\hline
$C_{BB}/\Lambda^{4}$  &$[-0.0212;0.0218]$  &$[-0.0475;0.0465]$ &$[-0.0173;0.0221]$ \\
                      
$C_{BW}/\Lambda^{4}$  &$[-0.0514;0.0518]$  &$[-0.1276;0.1271]$ &$[-0.0455;0.0456]$ \\
                      
$C_{\widetilde{B}W}/\Lambda^{4}$   &$[-0.0226;0.0222]$  &$[-0.0263;0.0321]$ &$[-0.0129;0.0179]$ \\
                      
$C_{WW}/\Lambda^{4}$   &$[-0.0678;0.0702]$  &$[-0.0678;0.0683]$ &$[-0.1108;0.1110]$ \\
\end{tabular}
\end{ruledtabular}
\end{table}

Our most sensitive limits in Eqs.~(\ref{eq.17})-(\ref{eq.20}) can be compared to phenomenological studies for the processes $pp\rightarrow ZZ\rightarrow 4\ell$ \cite{Yilmaz:2020ser} and $pp\rightarrow ZZ\rightarrow 2\ell2\nu$ \cite{Yilmaz:2021dbm} in the FCC-hh collider with 100 TeV. Although the FCC-hh collider has very high center-of-mass energy compared to the CLIC, the obtained limits are 4-8 times more sensitive than Ref.~\cite{Yilmaz:2020ser}. In addition, when the limits in Ref.~\cite{Yilmaz:2021dbm} are examined, it is seen that the limits of $C_{BB}$, $C_{BW}$ and $C_{\widetilde{B}W}$ couplings are better in our paper.

\section{Conclusions}

The LHC offers exciting possibilities for new discoveries and important insights at high-energy scale. However, it is widely recognized that linear colliders, with their clean experimental conditions, are the ideal environment for exploring new physics beyond the SM. The ability to use polarized electron beams in linear colliders enhances the physics program, allowing for precise testing of the SM and exploring new physics beyond the SM. The CLIC, with its multi-TeV energy and multiple electron polarization options, is a particularly promising linear collider for these studies. In this study, we investigate the process $e^+e^- \to ZZ$ using unpolarized and polarized beams to search for effects of anomalous $ZZZ$ and $ZZ\gamma$ couplings at the CLIC. To investigate the process $e^+e^- \to ZZ$ at the CLIC with a center-of-mass energy of 3 TeV, a cut-based method is performed with the detector response using Delphes card. The final state particles consists of two charged leptons and pair neutrinos coming from $Z$ bosons, and the $e^{-}e^{+}\to \ell \ell \nu \nu$ background are also analyzed. The results show that the transverse momentum $p^\ell_T$ and pseudo-rapidity $\eta^{\ell}$ of the charged leptons, $p_T$ balance, missing energy transverse $E_T^{miss}$, angular separation $\Delta{R}({\ell_{1},\ell_{2}})$ and invariant mass of the final state charged leptons $M_{\ell \ell}$ are the key parameters in differentiating the signal from the background.
After applying the selected cuts, the sensitivities of each anomalous coupling are obtained via 
$\chi^2$ method using the transverse momentum of the dilepton system at CLIC for unpolarized and polarized beams with $\sqrt{s}$=3 TeV. The findings of this study show that the CLIC can improve the sensitivity limits on $C_{\widetilde{B}W}/{\Lambda^4}$, $C_{BB}/{\Lambda^4}$, $C_{BW}/{\Lambda^4}$ and $C_{WW}/{\Lambda^4}$ parameters defining the aNTGC $ZZZ$ and $ZZ\gamma$ with respect to the experimental limits of LHC.

\section{Data Availability Statement}

This manuscript has no associated data or the data will not be deposited.


\begin{thebibliography}{99}

\bibitem{Aaboud:2019bza}
M.~Aaboud {\it et al.} [ATLAS Collaboration],
{\it JHEP} {\bf 10}, 127 (2019)
[arXiv:1905.07163 [hep-ex]].

\bibitem{Beyer:2006wsa}
M.~Beyer, W.~Kilian, P.~Krstonosic, K.~Monig, J.~Reuter, E.~Schmidt and H.~Schroder,
{\it Eur.\ Phys.\ J.\ C} {\bf 48}, 353-388 (2006)
[arXiv:hep-ph/0604048].

\bibitem{Fleper:2017tmq}
C.~Fleper, W.~Kilian, J.~Reuter and M.~Sekulla,
{\it Eur.\ Phys.\ J.\ C} {\bf 77}, 120 (2017)
[arXiv:1607.03030 [hep-ph]].

\bibitem{Demirci:2022acw}
M.~Demirci and A.~B.~Balantekin,
{\it Phys.\ Rev.\ D} {\bf 106}, 073003 (2022)
[arXiv:2209.13720 [hep-ph]].

\bibitem{Boland:2016yqz}
M.~J.~Boland {\it et al.} [CLIC and CLICdp Collaborations],
``Updated baseline for a staged Compact Linear Collider,'' 
(2016)
[arXiv:1608.07537 [physics.acc-ph]].

\bibitem{Roloff:2018tvb}
P.~Roloff, R.~Franceschini, U.~Schnoor and A.~Wulzer, 
``The Compact Linear $e^+e^-$ Collider (CLIC): Physics Potential,'' 
(2018)
[arXiv:1812.07986 [hep-ex]].

\bibitem{Fujii:2018ujn}
K.~Fujii {\it et al.} [The Linear Collider Collaboration],
``The role of positron polarization for the inital 250 GeV stage of the International Linear Collider,'' 
DESY 17-237 (2018)
[arXiv:1801.02840 [hep-ph]].

\bibitem{Shaposhnikovab:1987vbn}
M.~E.~Shaposhnikov,
{\it Nucl.\ Phys.\ B} {\bf 287}, 757-775 (1987).

\bibitem{Nelson:1992ded}
A.~E.~Nelson, D.~B.~Kaplan and A.~G.~Cohen,
{\it Nucl.\ Phys.\ B} {\bf 373}, 453-478 (1992).

\bibitem{Sakharov:1967hla}
A.~D.~Sakharov,
{\it Pisma.\ Zh.\ Eksp.\ Teor.\ Fiz.} {\bf 5}, 32-35 (1967) [Usp.\ Phys.\ Nauk {\bf 161}, 61-64 (1991)].

\bibitem{Degrande:2014ydn}
C.~Degrande,
{\it JHEP} {\bf 02}, 101 (2014)
[arXiv:1308.6323 [hep-ph]].

\bibitem{Gounaris:2000wlp}
G.~J.~Gounaris, J.~Layssac and F.~M.~Renard,
{\it Phys.\ Rev.\ D} {\bf 61}, 073013 (2000)
[arXiv:hep-ph/9910395].

\bibitem{Gounaris:2000jsx}
G.~J.~Gounaris, J.~Layssac and F.~M.~Renard,
{\it Phys.\ Rev.\ D} {\bf 62}, 073013 (2000)
[arXiv:hep-ph/0003143].

\bibitem{Choudhury:2001yvz}
D.~Choudhury, S.~Dutta, S.~Rakshit and S.~Rindani,
{\it Int.\ J.\ Mod.\ Phys.\ A} {\bf 16}, 4891-4910 (2001)
[arXiv:hep-ph/0011205].

\bibitem{Gounaris:2006edf}
G.~J.~Gounaris,
{\it Acta.\ Phys.\ Polon.\ B} {\bf 37}, 1111-1126 (2006)
[arXiv:hep-ph/0510061].

\bibitem{Dutta:2009tvq}
S.~Dutta, A.~Goyal and Mamta,
{\it Eur.\ Phys.\ J.\ C} {\bf 63}, 305-315 (2009)
[arXiv:0901.0260 [hep-ph]].

\bibitem{Chang:1995gwx}
D.~Chang, W-Y.~Keung and P.~B.~Pal,
{\it Phys.\ Rev.\ D} {\bf 51}, 1326 (1995)
[arXiv:hep-ph/9407294]].

\bibitem{Grzadkowski:2016evb}
B.~Grzadkowski, O.~M.~Ogreid and P.~Osland,
{\it JHEP} {\bf 05}, 025 (2016) [Erratum: {\it JHEP} {\bf 11}, 002 (2017)]
[arXiv:1603.01388 [hep-ph]].

\bibitem{Belusca:2018hhd}
H.~B{\'e}lusca-Maito, A.~Falkowski, D.~Fontes, J.~C.~Romao and J.~P.~Silva,
{\it JHEP} {\bf 04}, 002 (2018)
[arXiv:1710.05563 [hep-ph]].

\bibitem{Hagiwara:1987fqw}
K.~Hagiwara, R.~D.~Peccei, D.~Zeppenfeld and K.~Hikasa,
{\it Nucl.\ Phys.\ B} {\bf 282}, 253-307 (1987).

\bibitem{Rodriguez:2009rnw}
A.~Guti{\'e}rrez-Rodr{\'{\i}}guez, M.~A.~Hern{\'a}ndez-Ru{\'{\i}}z and M.~A.~P{\'e}rez,
{\it Phys.\ Rev.\ D} {\bf 80}, 017301 (2009)
[arXiv:0808.0945 [hep-ph]].

\bibitem{Ananthanarayan:2014cal}
B.~Ananthanarayan, J.~Lahiri, M.~Patra and S.~D.~Rindani,
{\it JHEP} {\bf 08}, 124 (2014)
[arXiv:1404.4845 [hep-ph]].

\bibitem{Rahaman:2016nzs}
R.~Rahaman and R.~K.~Singh,
{\it Eur.\ Phys.\ J.\ C} {\bf 76}, 539 (2016)
[arXiv:1604.06677 [hep-ph]].

\bibitem{Rahaman:2017qed}
R.~Rahaman and R.~K.~Singh,
{\it Eur.\ Phys.\ J.\ C} {\bf 77}, 521 (2017)
[arXiv:1703.06437 [hep-ph]].

\bibitem{Ellis:2020ekm}
J.~Ellis, S.~F.~Ge, H.~J.~He and R.~Q.~Xiao,
{\it Chin.\ Phys.\ C} {\bf 44}, 063106 (2020)
[arXiv:1902.06631 [hep-ph]].

\bibitem{Fu:2021jec}
Q.~Fu, J.~C.~Yang, C.~X.~Yue and Y.~C.~Guo,
{\it Nucl.\ Phys.\ B} {\bf 972}, 115543 (2021)
[arXiv:2102.03623 [hep-ph]].

\bibitem{Ellis:2021rop}
J.~Ellis, H.~J.~He and R.~Q.~Xiao,
{\it Sci.\ China\ Phys.\ Mech.\ Astron.} {\bf 64}, 221062 (2021)
[arXiv:2008.04298 [hep-ph]].

\bibitem{Spor:2022ssp}
S.~Spor, E.~Gurkanli and M.~K\"{o}ksal,
{\it Nucl.\ Phys.\ B} {\bf 979}, 115785 (2022)
[arXiv:2203.02352 [hep-ph]].

\bibitem{Yang:2022tgw}
J.~C.~Yang, Y.~C.~Guo and L.~H.~Cai,
Nucl.\ Phys.\ B {\bf 977}, 115735 (2022)
[arXiv:2111.10543 [hep-ph]].

\bibitem{Jahedi:2023abc}
S.~Jahedi and J.~Lahiri,
JHEP {\bf 04}, 085 (2023)
[arXiv:2212.05121 [hep-ph]].

\bibitem{Senol:2018gvg}
A.~Senol, H.~Denizli, A.~Yilmaz, I.~T.~Cakir, K.~Y.~Oyulmaz, O.~Karadeniz and O.~Cakir,
{\it Nucl.\ Phys.\ B} {\bf 935}, 365-376 (2018)
[arXiv:1805.03475 [hep-ph]].

\bibitem{Rahaman:2019tnp}
R.~Rahaman and R.~K.~Singh,
{\it Nucl.\ Phys.\ B} {\bf 948}, 114754 (2019)
[arXiv:1810.11657 [hep-ph]].

\bibitem{Senol:2019ybv}
A.~Senol, H.~Denizli, A.~Yilmaz, I.~T.~Cakir and O.~Cakir,
{\it Acta\ Phys.\ Pol.\ B} {\bf 50}, 1597 (2019)
[arXiv:1906.04589 [hep-ph]].

\bibitem{Senol:2020hbh}
A.~Senol, H.~Denizli, A.~Yilmaz, I.~T.~Cakir and O.~Cakir,
{\it Phys.\ Lett.\ B} {\bf 802}, 135255 (2020)
[arXiv:1910.03843 [hep-ph]].

\bibitem{Yilmaz:2020ser}
A.~Yilmaz, A.~Senol, H.~Denizli, I.~T.~Cakir and O.~Cakir,
{\it Eur.\ Phys.\ J.\ C} {\bf 80}, 173 (2020)
[arXiv:1906.03911 [hep-ph]].

\bibitem{Yilmaz:2021dbm}
A.~Yilmaz, 
{\it Nucl.\ Phys.\ B} {\bf 969}, 115471 (2021)
[arXiv:2102.01989 [hep-ph]].

\bibitem{Hernandez:2021wsz}
A.~I.~Hern{\'a}ndez-Ju{\'a}rez, A.~Moyotl and G.~Tavares-Velasco,
{\it Eur.\ Phys.\ J.\ C} {\bf 81}, 304 (2021)
[arXiv:2102.02197 [hep-ph]].

\bibitem{Lombardi:2022tgv}
D.~Lombardi, M.~Wiesemann and G.~Zanderighi,
{\it Phys.\ Lett.\ B} {\bf 824}, 136846 (2022)
[arXiv:2108.11315 [hep-ph]].

\bibitem{Senol:2022psb}
A.~Senol, S.~Spor, E.~Gurkanli, V.~Cetinkaya, H.~Denizli and M.~K\"{o}ksal,
{\it Eur.\ Phys.\ J.\ Plus} {\bf 137}, 1354 (2022)
[arXiv:2205.02912 [hep-ph]].

\bibitem{Spor:2023ywc}
S.~Spor,
{\it Nucl.\ Phys.\ B} {\bf 991}, 116198 (2023)
[arXiv:2207.11585 [hep-ph]].

\bibitem{Aaboud:2018ybz}
M.~Aaboud {\it et al.} [ATLAS Collaboration],
{\it JHEP} {\bf 12}, 010 (2018)
[arXiv:1810.04995 [hep-ex]].

\bibitem{Aaboud:2018onm}
M.~Aaboud {\it et al.} [ATLAS Collaboration],
{\it Phys.\ Rev.\ D} {\bf 97}, 032005 (2018)
[arXiv:1709.07703 [hep-ex]].

\bibitem{Sirunyan:2021edk}
A.~M.~Sirunyan {\it et al.} [CMS Collaboration],
{\it Eur.\ Phys.\ J.\ C} {\bf 81}, 200 (2021)
[arXiv:2009.01186 [hep-ex]].

\bibitem{Alwall:2014cvc}
J.~Alwall, R.~Frederix, S.~Frixione, V.~Hirschi, F.~Maltoni, O.~Mattelaer, H.~S.~Shao, T.~Stelzer, P.~Torrielli and M.~Zaro, 
{\it JHEP} {\bf 07}, 079 (2014)
[arXiv:1405.0301 [hep-ph]].

\bibitem{Favereau:2014qaz}
J.~D.~Favereau, C.~Delaere, P.~Demin, A.~Giammanco, V.~Lemaître, A.~Mertens and M.~Selvaggi,
{\it JHEP} {\bf 02}, 057 (2014)
[arXiv:1307.6346 [hep-ex]].

\bibitem{Bierlich:2022tge}
C.~Bierlich {\it et al.},
{\it Sci.\ Post.\ Phys.\ Codeb.} {\bf 8}, 1 (2022)
[arXiv:2203.11601 [hep-ph]].

\bibitem{ExRootAnalysis}
http://madgraph.physics.illinois.edu/Downloads/ExRootAnalysis/

\bibitem{Brun:1997cvb}
R.~Brun and F.~Rademakers, 
{\it Nucl.\ Instrum.\ Meth.\ A} {\bf 389}, 81-86 (1997).

\end{thebibliography}
\end{document}